\documentclass[aps,pre,twocolumn,groupedaddress,floatfix]{revtex4}

\usepackage{amsmath,amssymb,amsfonts,bm,xcolor,graphicx}

\DeclareGraphicsExtensions{.eps}

\newcommand{\vct}[1]{{\ensuremath{\bm{#1}}}}

\newcommand{\pfc}[2][]{\ensuremath{\widetilde{#2}_{#1}}}
\newcommand{\nfc}[2][]{\ensuremath{#2}_{#1}}
\newcommand{\mol}[1]{\ensuremath{#1\,\text{mol\,l}^{-1}}}

\begin{document}

\title{Models of electrolyte solutions from molecular descriptions: The example of NaCl solutions}

\author{John Jairo \surname{Molina}$^{1,2,3}$}
\email{john.molina@etu.upmc.fr}
\author{Jean-Fran\c cois \surname{Dufr\^eche}$^{1,2,3}$}
\email{jean-francois.dufreche@upmc.fr}
\author{Mathieu Salanne$^{1,2}$}
\author{Olivier Bernard$^{1,2}$}
\author{Marie Jardat$^{1,2}$}
\author{Pierre Turq$^{1,2}$}
\affiliation{$^1$ UPMC-Universit\'e Paris 06, UMR 7195, PECSA, F-75005 Paris, France \\
$^2$ CNRS, UMR 7195, PECSA, F-75005 Paris, France \\
$^3$ Institut de Chimie S\'eparative de Marcoule (ICSM),
UMR 5257 CEA--CNRS--Universit\'e~Montpellier~2,
Site de Marcoule, B\^atiment 426, BP 17171, 30207 Bagnols-sur-C\`eze Cedex, France
}


\begin{abstract}
We present a method to derive implicit solvent models of electrolyte solutions from all-atom descriptions; providing analytical expressions of the thermodynamic and structural properties of the ions consistent with the underlying explicit solvent representation. Effective potentials between ions in solution are calculated to perform perturbation theory calculations, in order to derive the best possible description in terms of charged hard spheres. Applying this method to NaCl solutions yields excellent agreement with the all-atom model, provided ion association is taken into account.
\end{abstract}


\maketitle
Since the pioneering works of Debye, H\"uckel, and Onsager, electrolyte solutions have been commonly described by continuous solvent models, for which the McMillan-Mayer theory~\cite{McMillan45} provides a rigorous statistical-mechanical foundation.
Within that level of description, simple phenomenological models such as the primitive model (PM), for which the ions are assimilated to charged hard spheres~\cite{Barthel}, can lead to explicit formulas for the thermodynamic and structural properties (e.g., with the help of the mean spherical approximation (MSA)~\cite{Blum1} or the binding MSA (BIMSA)~\cite{Blum95}).
These models are the most practical to use~\cite{Dufreche05}, since they allow for a direct link between the experimental measurements and the microscopic parameters of the system. Nevertheless, they ignore the molecular structure of the solvent. Consequently, they cannot properly account for the complex specific effects of the ions, which appear in numerous biological, chemical, and physical interfacial phenomena~\cite{Jungwirth06,Kunz04}, without further developments.

An alternative procedure consists in carrying out molecular simulations, where both the solvent and solute are treated explicitly. After a rigorous averaging over the solvent configurations, a coarse-grained description of the ions, which still includes the effect of the solvent structure, can be obtained~\cite{Hess06bis,Kalcher09,Gavryushov06,Lyubartsev95}. However, this set of methods is purely numeric; they do not provide any analytical expression for thermodynamic quantities. They are therefore restricted to simple geometries~\cite{Horinek07,Lund08} (bulk solutions or planar interfaces). The description of complex systems, such as porous or electrochemical materials, is still based on continuous solvent models~\cite{VanDamme09}.

In this letter we present a method aimed at bridging the gap between analytical and numerical approaches. It is based on the application of liquid perturbation theory (LPT)~\cite{Hansen} to effective ion-ion potentials extracted from molecular dynamics (MD) results. Different approximations of the PM are employed for the case of NaCl electrolyte solutions: a two component model (MSA2), that only takes free ions into account, and two different three component models (MSA3 and BIMSA3), which include a third species (the contact ion pair). As we proceed to show, LPT allows us to select the best simple model which accurately accounts for the thermodynamics and the physical-chemistry of the system.

The first stage consists in calculating the McMillan-Mayer effective ion-ion interaction potentials $V_{ij}^{\text{eff}}(r)$, by inverting the radial distribution functions (RDF) $g_{ij}(r)$ obtained by MD. The simulations were carried out on a box of 2000 water molecules and 48 NaCl pairs using the same interaction potentials as in reference~\cite{Rasaiah01}. This setup corresponds to a concentration of \mol{0.64}. NPT ensemble sampling at standard pressure and temperature was enforced, with a time step of 1~fs and a pressure bath coupling constant of 1~ps. An equilibration run of 0.25~ns was followed by a production run of 0.6~ns for five different initial configurations. The averages of the resulting RDF were then used for the potential inversion via the HNC closure~\cite{Hansen}. These effective potentials are assumed to be concentration independent and will be used for simulations at all concentrations.

\begin{figure}
  \includegraphics[scale=0.30]{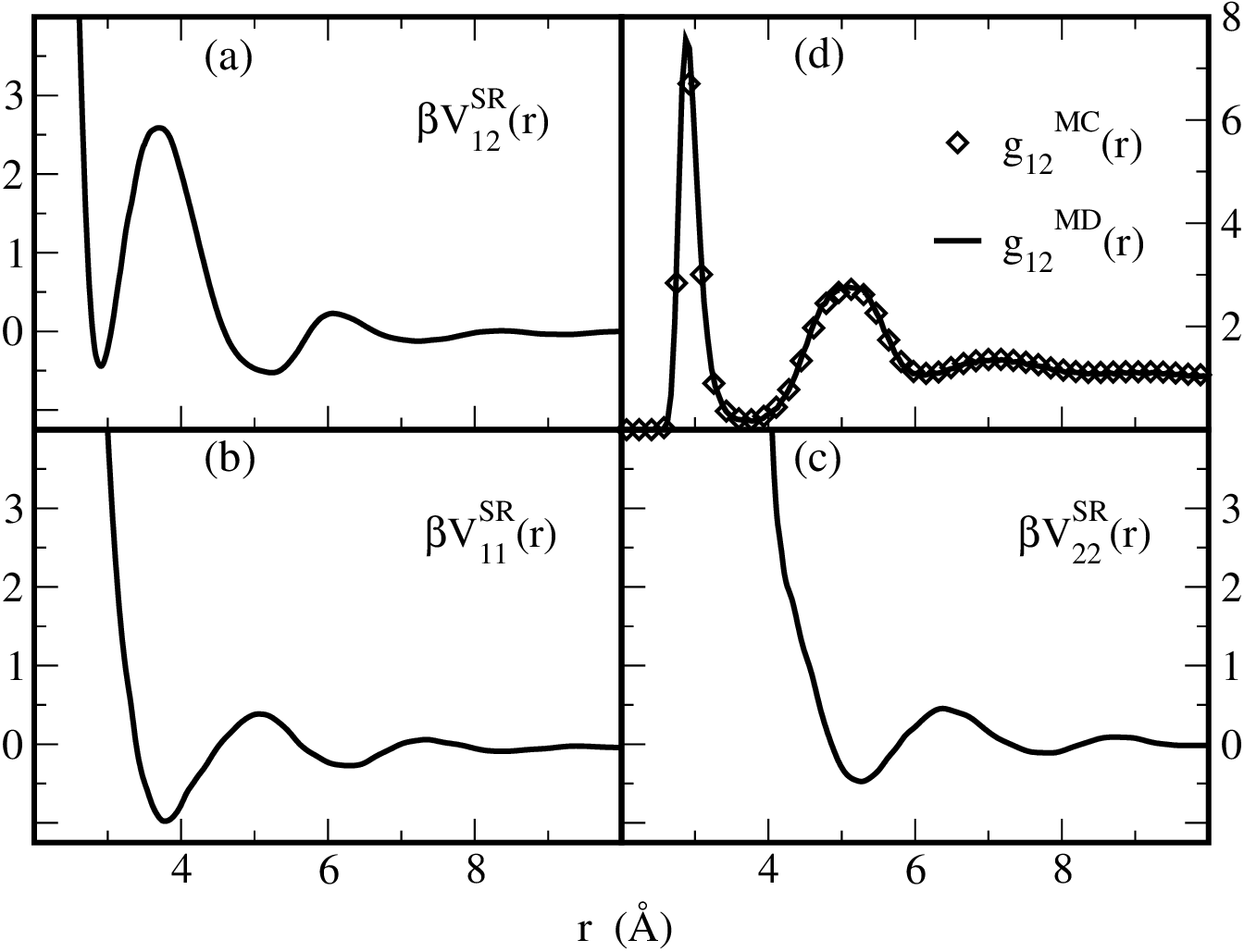}
  \caption{\label{fig:1}Effective McMillan-Mayer short-range pair potentials extracted from explicit solvent simulations using the HNC closure. (a) Cation anion, (b) cation cation, (c) anion anion, (d) cation anion RDF obtained from explicit solvent MD and implicit solvent MC simulations.}
\end{figure}

Subtracting the long-range Coulombic potential $V^{\text{LR}}_{ij}(r)$ (which depends on the dielectric constant of the solvent) from $V^{\text{eff}}_{ij}(r)$, we obtain the short-range contribution $V^{\text{SR}}_{ij}(r)$ to the effective potentials. These are given in Fig.~\ref{fig:1} (species 1 and 2 refer to Na$^+$ and Cl$^-$ free ions, respectively). All the short-range potentials exhibit oscillations corresponding to the solvent layering between the ions, but this effect is particularly important for the cation-anion interaction: a considerable potential barrier ($\gtrsim 2k_{\text{B}}T$) separates the first two attractive wells. To serve as a reference, Monte Carlo (MC) simulations were performed with these effective potentials; a comparison between MD and MC RDF is also provided in Fig.~\ref{fig:1}. The excellent agreement between both sets of RDF validates the HNC inversion procedure~\cite{Lyubartsev02}, and allows us to compute all ion thermodynamic properties through implicit solvent MC simulations.  

The second stage of our coarse-graining procedure consists in applying LPT, in order to deduce the best analytical model of electrolyte solutions which reproduces this molecular description. The principle of LPT is to describe the properties of a given system in terms of those of a well known reference system, with the difference between them treated as a perturbation in the reference potential. Assuming pairwise additive potentials, $V_{ij} = V_{ij}^{(0)} + \varDelta V_{ij}$, a first-order truncated  expression for the free energy density of the system $\beta f_v$ is obtained,
\begin{equation}
  \label{eqn:w1}
    \beta f_v \lesssim \beta f_v^{(0)} + \frac{1}{2}\beta\sum_{i,j}\rho_i\rho_j\int\text{d}\vct{r}\, g_{ij}^{(0)}(r) \varDelta V_{ij}(r)
\end{equation}
which depends only on the free-energy density $f_v^{(0)}$ and RDF $g^{(0)}$ of the reference fluid, with $\beta = (k_{\mathrm{B}}T)^{-1}$ and $\rho_i$ the concentration of species $i$. The Gibbs-Bogoliubov inequality~\cite{Hansen} ensures that the right-hand side of Eq.~(\ref{eqn:w1}) is actually a strict upper bound. Once a reference system has been chosen, the expression on the right-hand side of Eq.~(\ref{eqn:w1}) must be minimized with respect to the parameters defining the reference. This procedure yields the best first-order approximation to the free energy of the system under consideration. 

For a system of charged particles in solution, the natural reference is the PM, defined in terms of the charge and diameter ($\sigma_i$) of each species. In this case, the perturbing potentials are just the short-range effective potentials computed above ($\Delta V_{ij} = V^{\text{SR}}_{ij}$). We use the MSA~\cite{Blum1} solution to the PM, since it provides analytical expressions for both the free energy and the RDF. The perturbation term is evaluated using an exponential approximation to the RDF obtained within the MSA, $g(r)=\exp{\left[g_{\rm MSA}(r)-1\right]}$, which removes any unphysical negative regions and improves the comparison with HNC calculations.

\begin{figure}
  \includegraphics[scale=0.30]{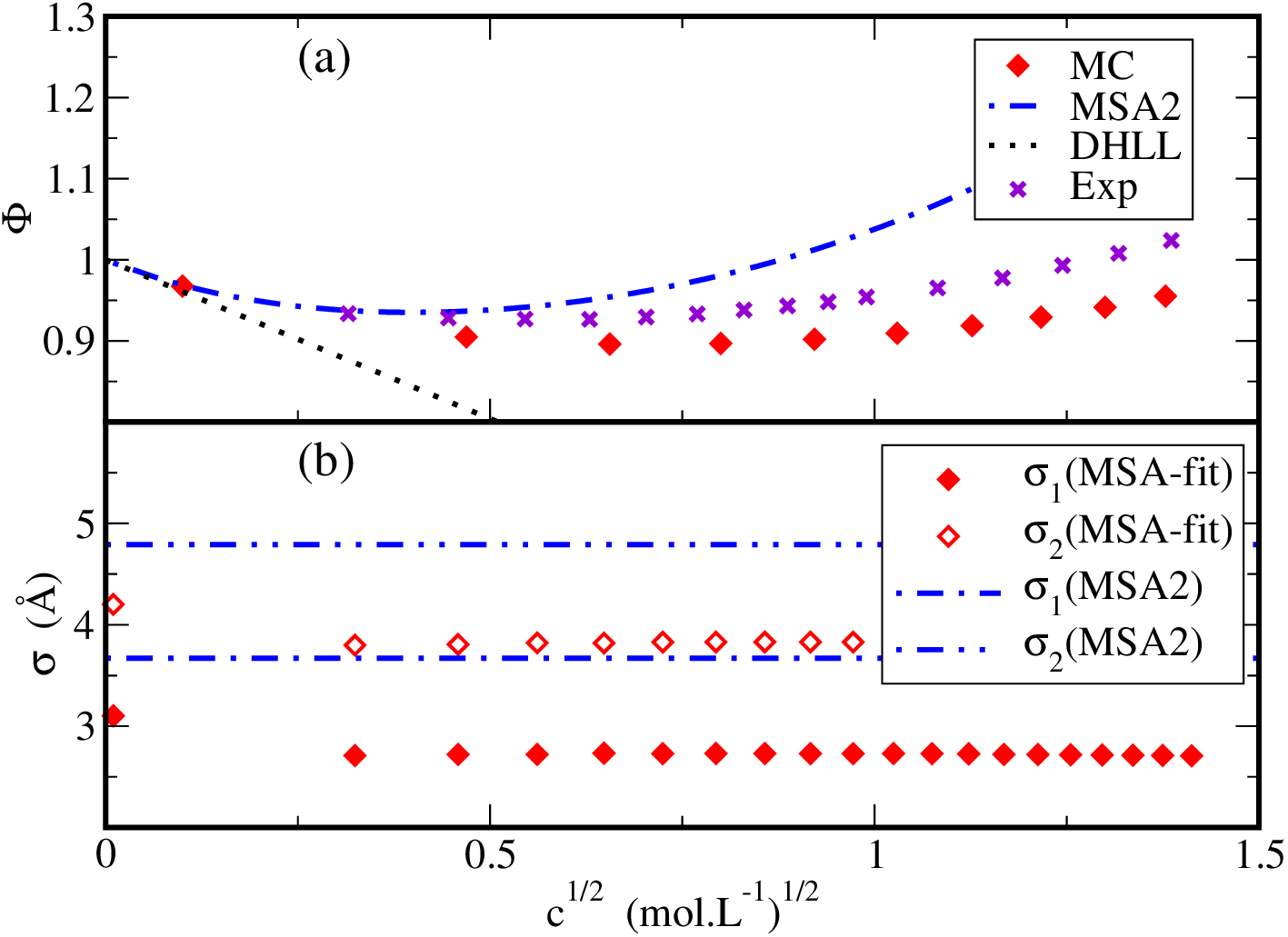}
  \caption{\label{fig:2}(Color online) (a) Osmotic coefficient $\Phi$ in the McMillan-Mayer frame of reference. (diamond) MC simulations, (dot dashed) MSA2, (dot) Debye H\"uckel Limiting law (DHLL), (cross) experiments (Ref.~\cite{Lobo} with the McMillan-Mayer to Lewis Randall conversion).
 (b) Minimization diameters. (dot dashed) MSA2 and (diamond) MSA-fit.}
\end{figure}

We first used LPT for a two-component system (Na$^+$ and Cl$^-$ free ions) within the MSA (model MSA2), for concentrations ranging from 0.1 to \mol{2.0}. The minimization leads to almost constant diameters on the whole range of concentration: $\sigma_1=3.67$~\AA \ and $\sigma_2=4.78$~\AA. As shown in Fig.~\ref{fig:2}, these parameters yield osmotic coefficients close to MC calculations only at very low concentration, i.e., $c\leq \mol{0.1}$ (experimental values are given for indicative purposes only, since a \emph{perfect} model will exactly match the MC results). For molar solutions, the LPT results differ considerably from MC calculations. This discrepancy can easily be understood by comparing the diameters found within the MSA2 calculation with the effective potentials given in Fig.~\ref{fig:1}. The anion/cation contact distance obtained within the MSA2 calculation is $4.2$~\AA, which is in the region of the second minimum of the effective potential and corresponds to the situation where there is a single layer of water molecules between the ions. The first minimum of the potential, which corresponds to the contact ion pair (CIP) is thus completely ignored by the MSA2 calculation. If the MSA diameters are directly fitted to reproduce the MC osmotic pressure, much smaller values are obtained. These MSA-fit hydrated diameters, which are compared to the MSA2 diameters in the bottom part of Fig.~\ref{fig:2}, are averages of the CIP and the solvent-separated ion pair. 

\begin{figure}
  \includegraphics[scale=0.30]{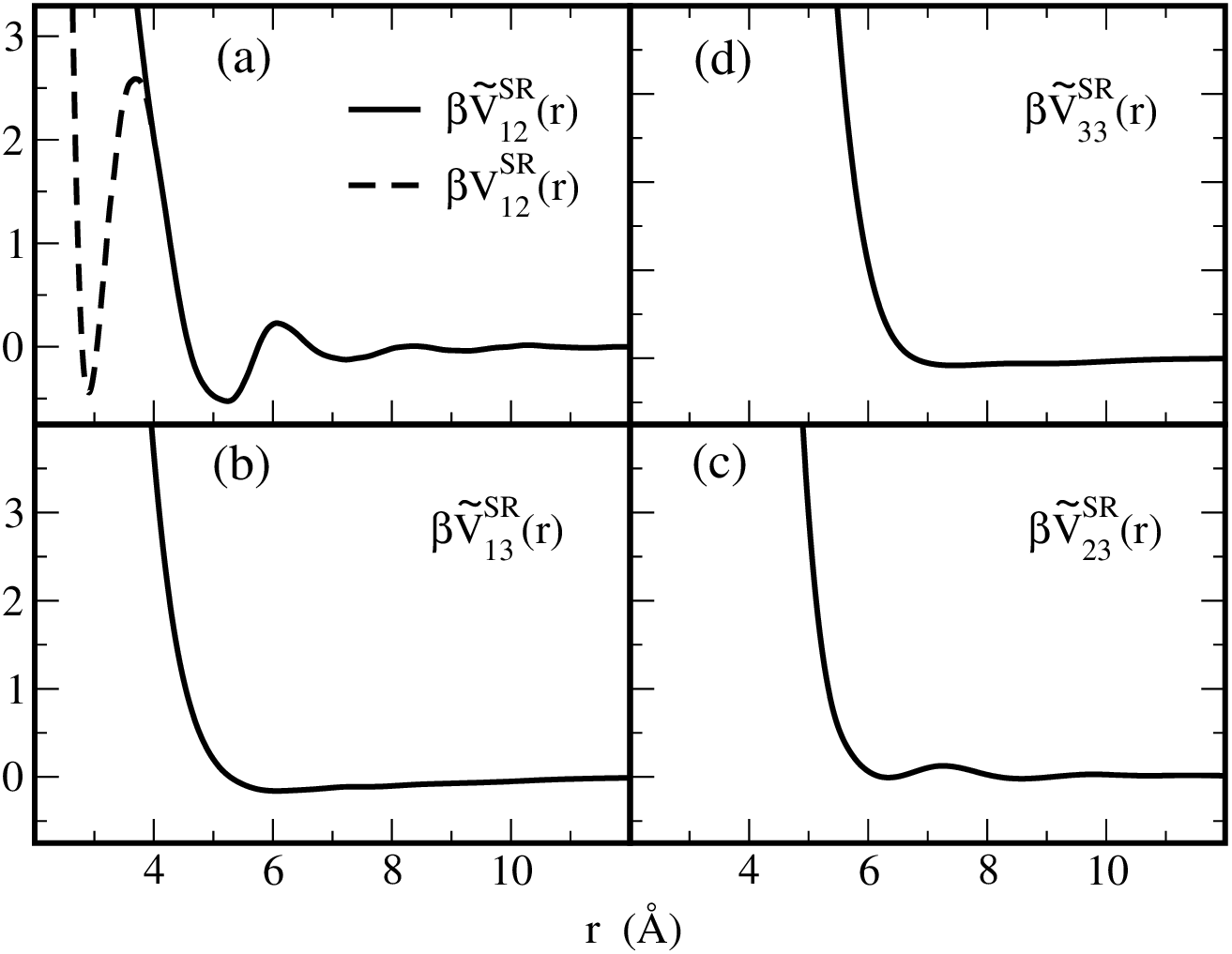}
  \caption{\label{fig:3} Effective pair potentials derived for MSA3 and BIMSA3. (a) Cation anion (dashed line: without taking the pair into account), (b) pair cation, (c) pair anion, and (d) pair pair. The internal potential of the pair $\beta\pfc[\text{int}]{V}(r)$ is set equal to $\beta V^{\text{eff}}_{ij}(r)$ for distances less than 4~\AA.}
\end{figure}

To overcome this difficulty, we have explicitly introduced the CIP in our model (species 3). Straightforward calculations, based on a characteristic-function formalism, allow us to define an equivalent model in which the free ions and the CIP are explicitly taken into account~\cite{Ciccotti84,Dufreche03bis}. We apply this formalism by defining a pair as an anion and a cation at a distance less than 4~\AA, which corresponds to the position of the effective potential maximum. The interaction between free, like charges in this new system remains unchanged, and the cation-anion interactions are easily approximated by extrapolating the original potential at the barrier separating pairs from free ions (as shown in Fig.~\ref{fig:3}). We assume that the interaction potential is averaged over the rotational degrees of freedom of the CIP and thus pairwise additive.  Hereafter, the quantities referring to such a three-component model are written with a tilda symbol. 
The short-range potentials involving the pair can be derived, in the infinite dilution limit, from an average of the contributing ion interactions. In Fourier space, 
\begin{subequations}\label{eqn:pot}
  \begin{align}
    \pfc[3i]{V}^{\text{SR}}(\vct{k}) &= \pfc{w}(\vct{k}/2)\bigl[\nfc[1i]{V}^{\text{SR}} + \nfc[2i]{V}^{\text{SR}}\bigr](\vct{k}),\quad i=1,2 \\
    \pfc[33]{V}^{\text{SR}}(\vct{k}) &= \pfc{w}(\vct{k}/2)^2\bigl[\nfc[11]{V}^{\text{SR}} + \nfc[22]{V}^{\text{SR}} + 2\nfc[12]{V}^{\text{SR}}\bigr](\vct{k})
    \intertext{where $\pfc{w}(\vct{r})$ is the pair probability distribution}
    \pfc{w}(\vct{r}) &= {K}_{0}^{-1}e^{-\beta \pfc[\text{int}]{V}(r)}
\end{align}
\end{subequations}
$\pfc[\text{int}]{V}(r)$ is the internal part of the pair potential (see Fig.~\ref{fig:3}), and $K_0$ is the association constant, defined as:
\begin{equation}
{K}_0 = \int_0^\infty\text{d}r\, 4\pi r^2 e^{-\beta \pfc[\text{int}]{V}(r)} \label{eqn:k0} = 0.43~\text{L}.\text{mol}^{-1}
\end{equation}

The excess free-energy density of the original system $\beta\nfc[v]{f}^{\text{ex}}$ is that of the three component mixture $\beta\pfc[v]{f}^{\text{ex}}$ plus a correction term 
\begin{equation}
\beta\nfc[v]{f}^{\text{ex}} = \beta\pfc[v]{f}^{\text{ex}} - \pfc[3]{\rho}\ln{K}_0 \label{eqn:free},
\end{equation}
which is due to the change in standard chemical potential between the two component and three component models. It should be noted that the fraction of pairs is now an additional parameter in the minimization scheme, which serves to ensure chemical equilibrium. Within this representation, the pair can be modeled as a hard sphere (MSA3) or as a dumbbell-like CIP (BIMSA3)~\cite{Blum95}. Since we have no additional information, we consider only symmetric dumbbells. Furthermore, since analytic expressions for the RDF within BIMSA are not known, we approximate the dumbbell as a hard sphere when computing the perturbation term (this is not necessary for the reference term, since an expression for the free energy is available). Let $\pfc[c]{\sigma}$ be the diameter of the cation (anion) within the dumbbell, the diameter of the hard sphere representing this dumbbell is taken to be $\pfc[3]{\sigma}=\frac{4\sqrt{2}}{\pi}\pfc[c]{\sigma}$\footnote{The average contact distance between a symmetric dumbbell and an infinite plane at $\beta=0$.}. 

\begin{figure}
  \includegraphics[scale=0.30]{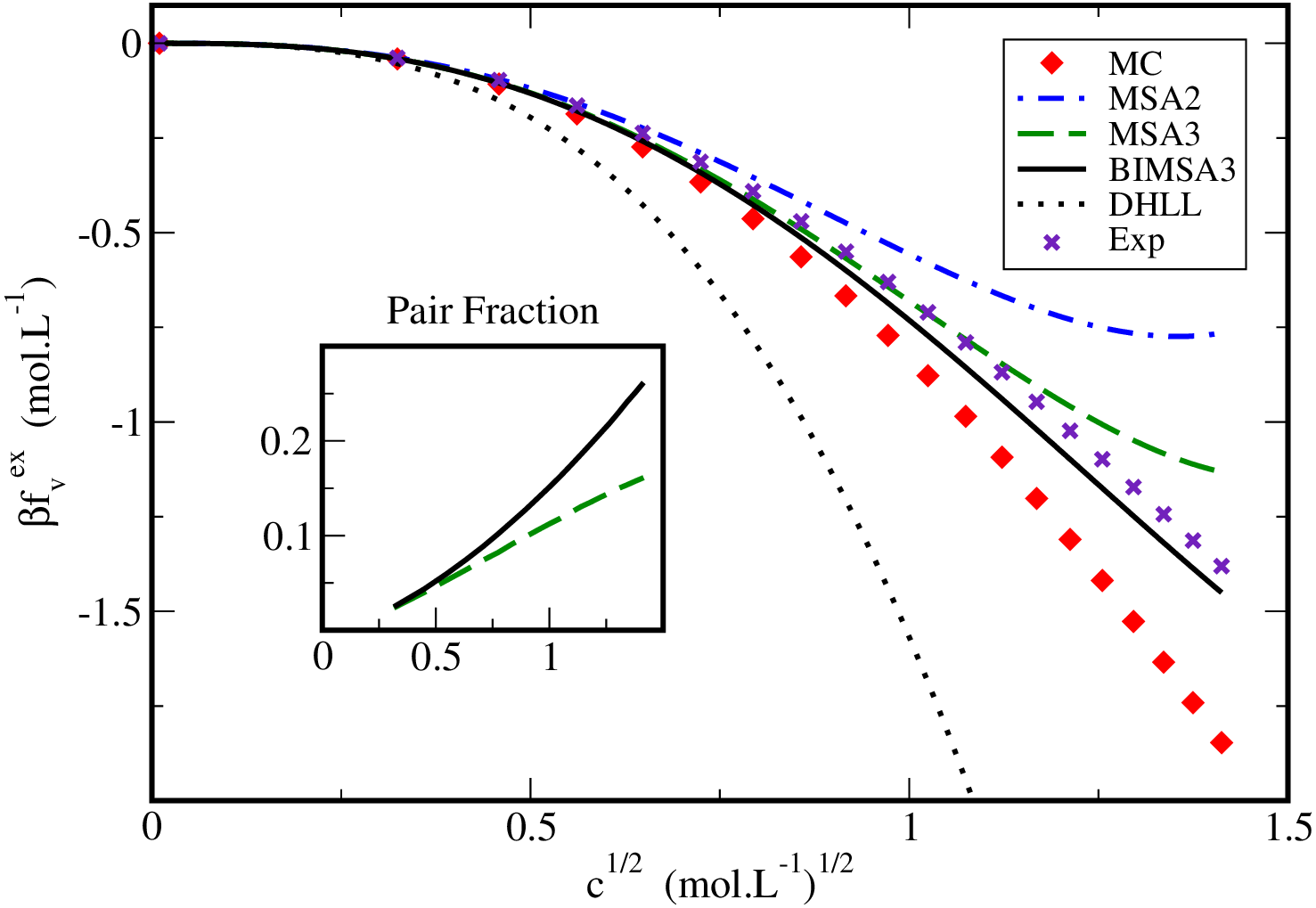}
  \caption{\label{fig:4}(Color online) Excess free-energy density $\beta f^{\text{ex}}_v$ as a function of the square root of the concentration $\sqrt{c}$. (diamond) MC simulations, (dot dashed) MSA2, (dashed) MSA3, (solid) BIMSA3, (dot) DHLL, and (cross) experiments. The inset gives the fraction of pairs (MSA3, BIMSA3) as a function of $\sqrt{c}$.}
\end{figure}

Using these two reference systems, the three-component MSA3 and BIMSA3, we obtain results in much better agreement with the MC simulations, as shown in Fig.~\ref{fig:4}. The diameters obtained for species 1, 2, and 3 are 3.65, 4.79, and 5.76~\AA~ for MSA3 and 3.69, 4.75 and 6.19~\AA~ for BIMSA3. The free ion diameters are similar for MSA2, MSA3, and BIMSA3. The pair diameter is smaller when modeled as a hard sphere (MSA3) than when modeled as a dumbbell (BIMSA3). At high concentration (about \mol{1}), the MSA3 overestimates the free energy, because the excluded volume repulsion becomes too important for the pairs to be represented as hard spheres. The BIMSA3 model is the closest to the MC simulation results. It is worth noting that even at the lowest concentration considered, the fraction of pairs (shown in the insert of Fig.~\ref{fig:4}), although less then 5\%, has a non-negligible effect on the thermodynamics of the system.
\begin{figure}
  \includegraphics[scale=0.30]{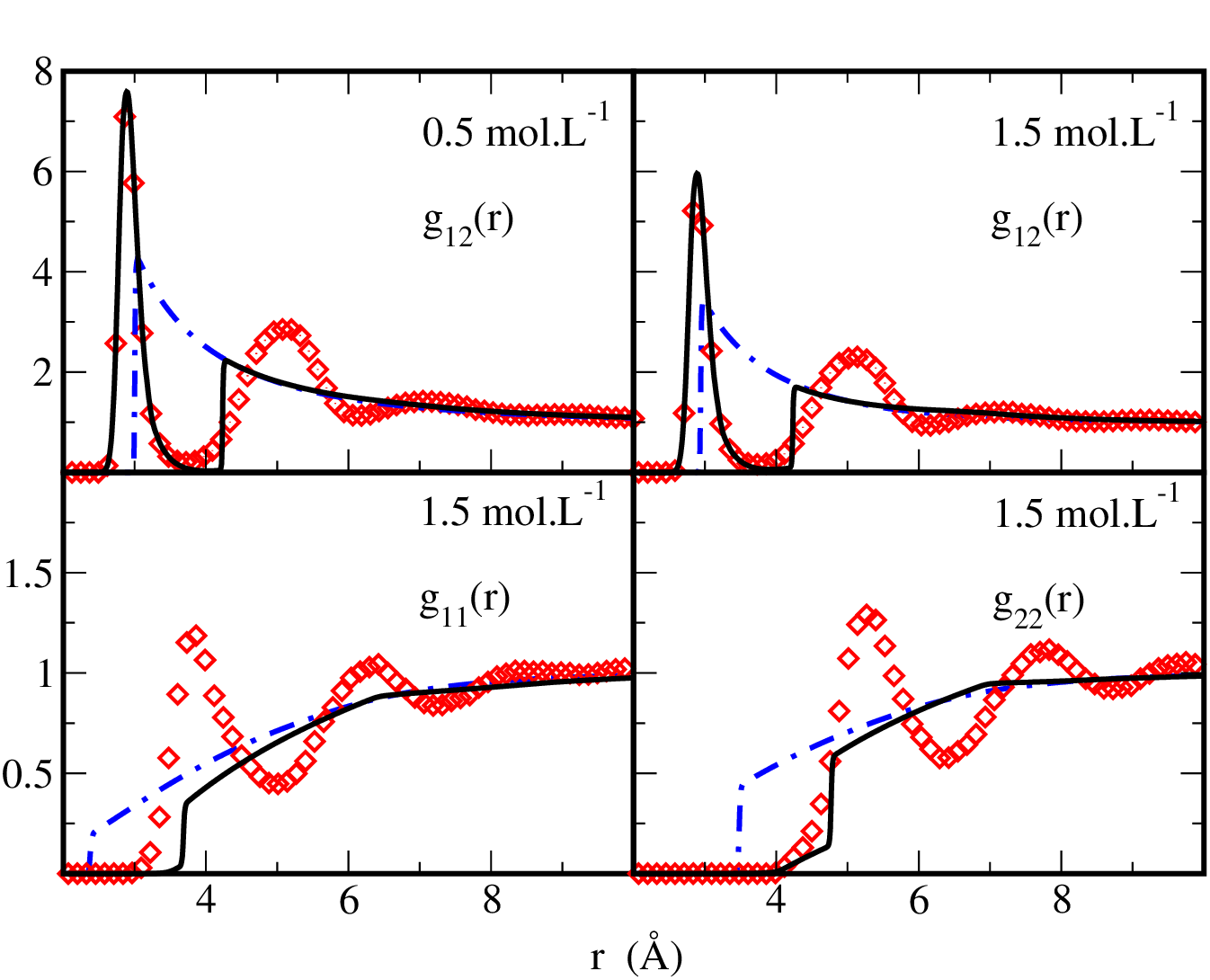}
  \caption{\label{fig:5}(Color online) RDF obtained from MC simulations (diamond), BIMSA3 (solid line), and MSA-fit (dot dashed) at two concentrations.}
\end{figure}

This procedure also provides an accurate description of the structure over the whole range of concentrations. A development similar to the one that leads to Eq.~(\ref{eqn:pot}) derives the average unpaired RDF from the corresponding paired quantities:
\begin{align}
  \rho_i\rho_j\nfc[ij]{g}(\vct{k}) &= \pfc[3]{\rho}\pfc{w}(\vct{k})\left(1-\delta_{ij}\right) +\pfc[i]{\rho}\pfc[j]{\rho}\pfc[ij]{g}(\vct{k})\notag\\
  &+ \pfc[3]{\rho}\pfc{w}(\vct{k}/2)\bigr[\pfc[i]{\rho}\pfc[3i]{g} + \pfc[j]{\rho}\pfc[3j]{g}\bigr](\vct{k})\\ 
  &+ \pfc[3]{\rho}^{\,2}\left[\pfc{w}(\vct{k}/2)\right]^2\pfc[33]{g}(\vct{k})\notag
\end{align}
 The RDF obtained within BIMSA3 are compared with the MC and MSA-fit results in Fig.~\ref{fig:5}.  Our BIMSA3 model accounts for the strong molecular peak of the CIP and provides the correct distances of minimal approach; whereas the naive MSA-fit procedure ignores the former and gives poor estimates for the latter. At larger separations, the BIMSA3 results do not reproduce the oscillations observed in the MC simulations, but the corresponding energy oscillations in the effective potentials are less than $k_{\mathrm{B}}T$.
In addition, the perturbation term of the BIMSA3 appears to be negligible compared to the reference term for concentrations less than \mol{1}. The perturbation can then be omitted to obtain a fully analytical theory, determined by the hard sphere diameters and the pair fraction given by LPT; with the free energy and the RDF given in terms of the BIMSA and MSA solutions, as described above. While the procedure we have followed uses two different approximations for the reference and perturbation terms (MSA vs BIMSA), these are known to be accurate for the systems under consideration and do not appear to be inconsistent with each other. 

To conclude, we have combined MD simulations with LPT to construct simple models of electrolyte solutions which account for the molecular nature of the solvent. The final result is fully analytical and it yields the thermodynamic and structural properties of the solution, in agreement with the original molecular description. The methodology can in principle be adapted to any molecular description of the system (MD simulations involving interaction potentials accounting for polarization effects or Car-Parrinello MD simulations for example) as long as the ion-ion RDF are known. It can also be generalized to study interfaces. The method appears to be a promising approach toward the description of the specific effects of ions, especially for complex systems whose modeling requires an analytic solution.

The authors are particularly grateful to Werner Kunz for fruitful discussions.

\end{document}